\documentclass[conference]{IEEEtran}
\IEEEoverridecommandlockouts
\usepackage{cite}
\usepackage{amsmath,amssymb,amsfonts}
\usepackage[utf8]{inputenc} 
\usepackage{algorithmic}
\usepackage{subfig}
\usepackage{algorithm}
\usepackage{graphicx}
\usepackage{multirow}
\usepackage{textcomp}
\usepackage{booktabs}
\usepackage{xcolor}

\usepackage{array}
\newcolumntype{L}[1]{>{\raggedright\let\newline\\\arraybackslash\hspace{0pt}}m{#1}}
\newcolumntype{C}[1]{>{\centering\let\newline\\\arraybackslash\hspace{0pt}}m{#1}}
\newcolumntype{R}[1]{>{\raggedleft\let\newline\\\arraybackslash\hspace{0pt}}m{#1}}

\definecolor{morange}{rgb}{0.8,0.2,0}
\definecolor{mblue}{rgb}{0,0.1,0.8}
\definecolor{mgreen}{rgb}{0,0.8,0.1}
\definecolor{mred}{rgb}{1,0,0}
\newcommand{\add}[1]{{\leavevmode\color{morange}{#1}}}

\begin{document}

\title{ Deep learning based intelligent IDS for Large-scale IoT networks\\

\author{
    \IEEEauthorblockN{
        Isha Andrade\IEEEauthorrefmark{1},
        Shalaka S. Mahadik\IEEEauthorrefmark{1},
        Mithun Mukherjee\IEEEauthorrefmark{1},
        Pranav M. Pawar\IEEEauthorrefmark{1},
        Raja Muthalagu\IEEEauthorrefmark{1}
    }
    \IEEEauthorblockA{\IEEEauthorrefmark{1}Department of Computer Science, Birla Institute of Technology and Science, Pilani, Dubai Campus,\\
    Dubai International Academic City, Dubai, United Arab Emirates\\
    Emails: \{f20200039, mithun, pranav, raja.m\}@dubai.bits-pilani.ac.in, mahadikshalaka4@gmail.com}
}

}

\maketitle

\maketitle

\begin{abstract}
The proliferation of large-scale IoT networks has been both a blessing and a curse. Not only has it revolutionized the way organizations operate by increasing the efficiency of automated procedures, but it has also simplified our daily lives.  However, while IoT networks have improved convenience and connectivity, they have also increased security risk due to unauthorized devices gaining access to these networks and exploiting existing weaknesses with specific attack types. The research proposes two lightweight deep learning (DL)-based intelligent intrusion detection systems (IDS). to enhance the security of IoT networks: the proposed convolutional neural network (CNN)-based IDS and the proposed long short-term memory (LSTM)-based IDS. The research evaluated the performance of both intelligent IDSs based on DL using the CICIoT2023 dataset. DL-based intelligent IDSs successfully identify and classify various cyber threats using binary, grouped, and multi-class classification. The proposed CNN-based IDS achieves an accuracy of 99.34\%, 99.02\% and 98.6\%, while the proposed LSTM-based IDS achieves an accuracy of 99.42\%, 99.13\%, and 98.68\% for binary, grouped, and multi-class classification, respectively.
 
\end{abstract}


\section{Introduction}
\begin{figure}[t]
\centerline{\includegraphics[height=8.5cm, width=9.5cm]{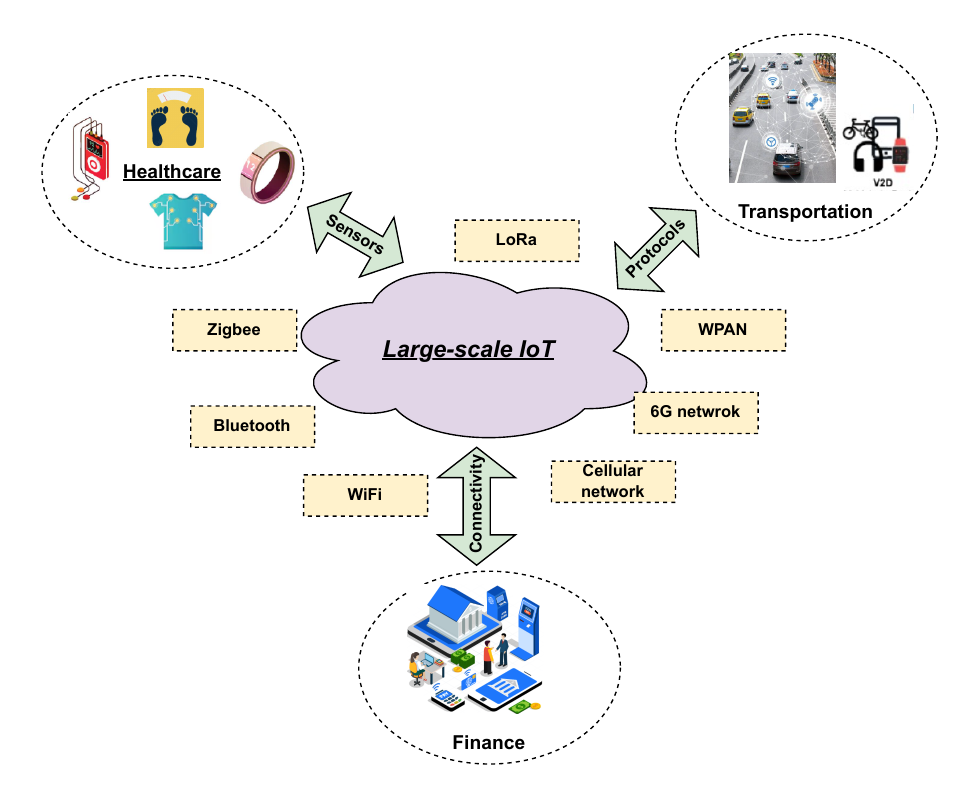}}
\caption{Example of large-scale IoT and its features}
\label{into-dia}
\end{figure}

The growing use of diverse large-scale Internet of Things (IoT) devices across various industries, including healthcare, finance, and transportation, is demonstrated in Fig. \ref{into-dia}. These large-scale IoT systems encompass various sensors, communication protocols, connectivity, and numerous other technologies, making cyber threats a paramount concern.  
The implementation of traditional security techniques is also challenging due to these unique features. Therefore, IoT devices can serve as gateways for cyberattacks, putting systems or applications on both individual and national scales \cite{intro1}.

There are various real-time examples available that highlight the importance of security mechanisms in today's IoT world. In March 2021, hackers infiltrated Verkada, a cloud-based video surveillance service, exposing private data and live feeds from more than 150,000 cameras \cite{intro2}. Moorfields Eye Hospital, one of the leading ophthalmology hospitals in the UAE, was targeted by a ransomware group \cite{intro3}. In 2024, the ransomware group targeted Roku, a TV streaming service provider, resulting in the compromise of approximately 576,000 accounts. It raised cybersecurity concerns for home-use IoT devices \cite{intro4}. Therefore, the development of a security model in the IoT context inspires the research.

\subsection{Motivation}
Recently, deep learning models such as CNN and LSTM have attracted many in academia for the development of efficient IDS \cite{ssm2, RW5}. The CNN models are capable of understanding features and extracting important information from raw data. The CNN model's various layers enable it to perform tasks such as object detection, cyber threat detection, and image identification, among others. They are more popular among academics as they can analyze network traffic effectively and detect vulnerabilities efficiently \cite{ssm1}. Similarly, LSTM models are good at retaining past data and predicting future data. These models have the ability to analyze network traffic logs or event sequences, identifying any sudden or drastic increases in network traffic that could potentially pose cyber threats. So, many researchers focused on it for the development of security mechanisms \cite{ssm2}. Therefore, the goal of the research is to develop a DL-based intelligent IDS using CNN and LSTM techniques.

\subsection{Contribution}
The primary contributions of the paper are outlined below:
\begin{itemize}

\item The research proposes two novel DL-based intelligent IDSs, i.e., the CNN-based IDS and the LSTM-based IDS. Both the proposed IDSs are lightweight in terms of layers and neurons used.

\item The proposed DL-based intelligent IDS employed the CICIoT2023 security dataset to detect and classify the malicious and benign network traffic. The research examined the performance of the proposed DL-based intelligent IDS using three classification types: binary (containing one attack and one benign class), grouped (containing seven attack classes and one benign class), and multi-class (containing 33 attack classes and one benign class). 

\item The research evaluated the performance of the proposed DL-based intelligent IDS in comparison to the state-of-the-art HetIoT CNN-IDS, focusing on binary, grouped, and multi-class classification perspectives.

\end{itemize}

The remainder of the research paper is organized as follows: Section II talks about the literature review. Section III outlines the research methodology, including datasets and their preprocessing steps, as well as DL-based intelligent IDS. Section IV presents the simulation setup and performance results, along with a summary. Section V concludes the research paper.

\section{Related work}

The section details various DL techniques used for the development of security models.

Paper \cite{RW5} investigates a federated learning (FL) technique for detecting large-scale IoT network assaults on the CICIoT2023 dataset. The paper utilises balance data and feature normalisation techniques to identify network attacks. The paper \cite{RW7} uses a DNN and a bidirectional LSTM (BiLSTM) model to learn nonlinear interactions and extract long-term dependencies in both directions. 
The model begins with an incremental principal component analysis (IPCA) approach to reduce the dimensionality of data. 

In the paper \cite{RW10}, the initial step is to convert the PCAP files into images, which are then preprocessed and categorized using a CNN. The CNN architecture included a convolutional layer with ReLU activation, a max pooling layer, and a fully connected dense layer. 
Next, paper \cite{RW11} describes a mixed deep learning model made up of an MLP and a CNN that can find distributed denial of service (DDoS) attacks in SDN settings. Shapley additive explaining (SHAP) and the Bayesian optimiser are used to select the best features and fine-tune hyperparameters, respectively. 
The paper \cite{RW13} discusses a hybrid lightweight DL model using a stacked autoencoder (SAE) and a CNN that is trained on the CICDDoS2019 dataset. 
An adaptive FLAD approach is proposed in the paper \cite{RW14} for detecting DDoS attacks. The FLAD methodology outperforms traditional FL approaches in both accuracy and time of convergence. 

A mixed LSTM and recurrent neural network model (LSTM-RNN) is suggested in paper \cite{RW15} as a way to find DDoS attacks in SDN networks. The model achieved an accuracy of 99.33\% for both the CICDDoS2017 and the CICDDoS2019 dataset. A multichain technology to detect DDoS attacks has been proposed in the paper \cite{RW16}. Existing methods for finding DDoS attacks at the application layer can only find a few types. To fix this, paper \cite{RW17} suggests an explicit duration RNN (EDRN) that is trained on the CICDDoS2019 dataset and works at the application layer. The proposed model achieves an accuracy, recall, and F1 score of 99.6\%, 99.3\%, and 99.2\%, respectively.

A deep CNN (DCNN) model is proposed in the paper \cite{RW18} to detect DDoS attacks in an SDN environment. The proposed DCNN model achieves an accuracy of 99.77\% for the CICDDoS2019 dataset. Four types of classifications are performed in paper \cite{RW19}are binary, multi-class with one-hot encoding, multi-class with a LabelEncoder, and multi-label classification. The CSE-CICIDS2018 dataset exclusively trained the proposed model, a four-layer MLP, on DDoS attacks. A DL-based contractive autoencoder is suggested in paper \cite{RW20} as a way to find DDoS attacks in three different datasets: CICIDS 2017, CICDDoS2019, and NSL-KDD. The performance is compared with a traditional autoencoder and other DL models. 

A hierarchical LSTM model is proposed in the paper \cite{RW21}. The model incorporates a dual-LSTM packet classifier to classify packet data of varying lengths and a single-LSTM session classifier to enhance accuracy by utilizing the maximum amount of session traffic. The paper \cite{RW22} presented a comparative study of various supervised (e.g., Alex Neural Network) and unsupervised models (e.g., Variational Autoencoder) trained on the CICDDoS2019 dataset. 

In paper \cite{RW23}, a new six-layer framework is suggested as a way to find attacks on IoT networks. The model is capable of identifying network attacks such as sinkholes, blackholes, DDoS attacks, etc. In paper \cite{RW24}, an SDN-IoT framework is employed to detect DDoS attacks, including zero-day attacks, using the CICDDoS2019 dataset. The proposed model consists of three LSTM layers along with one input and one output layer. 

Paper \cite{RW25} presented two hybrid DL models: a DCNNBiLSTM (a combination of a CNN and a BiLSTM model) and a DCNNGRU (a combination of a CNN and a gated recurrent unit). The performance of the proposed models was compared to DL models from previous research, such as CNN + LSTM, DNN, ResNet, etc. In paper \cite{RW26}, a stacked convolutional denoising AE (SCDAE) is use to get rid of the noise in the network traffic data. Next, a simple CNN and BiLSTM model are used to get the data's spatial and temporal features. BiLSTM is considered so the relationship between the traffic can be mined from both front and back. 

In paper \cite{RW27}, a DL-based IDS is proposed to detect cyber attacks from the CICDDoS2019 dataset. The proposed model is a combination of a BiLSTM, and BiGRU. A hybrid DL model of CNN, LSTM, AE, and DNN is introduced in the paper \cite{RW28} to detect DDoS attacks from the CICDDoS 2019 dataset. The proposed model achieves a detection rate of 71.42\%, a false detection rate of only 0.04\%, and, on average, an accuracy of 73.46\%.

\begin{table}[t]
 \footnotesize
  \centering
  \caption{Comparison of various state-of-the-art DL methods used for detecting IoT network attacks from three classifications perspectives}
  
    \begin{tabular}{L{1.05in}lll}
    \toprule
   
    & \multicolumn{3}{c}{Type of classifications} \\\cmidrule{2-4}
    & Binary & Grouped & Multi-Class \\
    \midrule
     FL + client DNN\cite{RW5} & 99\% & -     & - \\
    DNN-BiLSTM\cite{RW7}  & -     & -     & 93.13\% \\
    DNN\cite{RW10}   & 99.71\% & 98.76\% & 98.81\% \\
     MLP-CNN\cite{RW11} & -     & -     & 99.95\% \\
     DL + (ADASYN and SMOTE)\cite{RW12} & 99.97\% & -     & 99.99\% \\
     SAE-CNN-Detection (SCD)\cite{RW13} & 99.1\% & -     & 97.2\% \\
     LSTM-RNN\cite{RW15} & 99.33\% & -     & - \\
     EDRN\cite{RW17} & -     & -     & 99.6\% \\
     DCNN\cite{RW18}  & -     & -     & 99.77\% \\
     MLP\cite{RW19}   & 100\% & -     & 99.84\%, 99.7\% \\
     Contractive AE\cite{RW20} & -     & -     & 93.41\%, 97.58\% \\
     AlexNet, VAE\cite{RW22} & -     & -     & 98.81\%, 96.7\% \\
    FCFFN \cite{RW23} & 93.74\% & -     & - \\
     LSTM\cite{RW24}  & 99.8\% & -     & - \\
     DCNNBiLSTM, DCNNGRU\cite{RW25} & 99.95\%, 99.93\% & -     & - \\
     CNN-BiLSTM-Attention\cite{RW26} & -     & -     & 93.26\% \\
     BiLSTM-BiGRU\cite{RW27} & -     & -     & 99.77\% \\
     CNN-LSTM-DeepAE-DNN\cite{RW28} & -     & -     & 73.46\% \\
    \bottomrule
    \end{tabular}%
 \label{RW}%
\end{table}%

Table \ref{RW} summarizes various DL methods to detect cyberattacks in large-scale IoT networks. According to related work, the DL models presented in the literature take into account a variety of security datasets, including CICIDS2017, CICDDoS2019, NSL-KDD, and others. Neither of these datasets has been designed for IoT. The literature study shows that many of the DL models are hybrid such as a CNN-LSTM, CNN-BiLSTM and complex with a greater number of layers and neurons. Although these models have produced excellent results, they can produce a significant amount of computational overhead. So, the research, aims to propose two novel lightweight DL-based intelligent IDS, namely, the proposed CNN-based IDS and the proposed LSTM-based IDS. The research used the CICIoT2023 dataset, which is a recent benchmark dataset created exclusively for the IoT environment. The numerous models provided in the literature study lack focus on such specific IoT-related datasets, where research is concentrated. Next, section details the research methodology used in the research.

\section{Research methodology}
The section details the two novel DL models namely, proposed CNN-based IDS and LSTM-based IDS. Moreover the section details CICIoT23 dataset and data preprocessing steps. The section also highlights the dataset used for binary classification, grouped classification and multi-class classification.

\subsection{Dataset and data preprocessing}
\begin{figure}[t]
\centerline{\includegraphics[height=8cm, width=9.5cm]{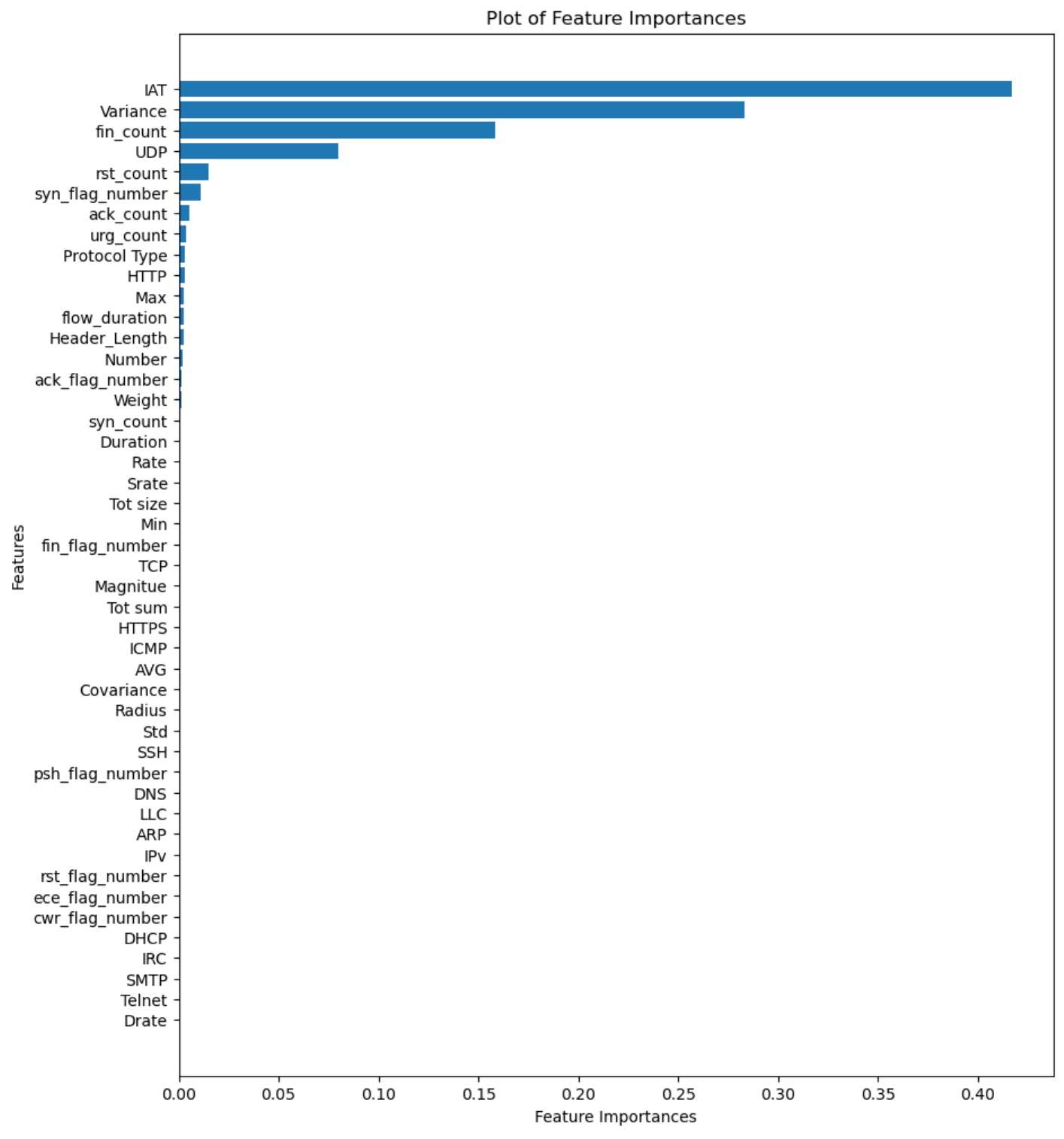}}
\caption{Feature importance graph \cite{isha1}}
\label{feat-imp}
\end{figure}

The CICIoT23 dataset is a publicly available security dataset provided by \cite{cic23dataset1, cic23dataset2}. The research used the CICIoT2023 dataset to classify IoT network attacks based on three classifications : binary, grouped, and multi-class. The original dataset consists of 168 network traffic CSV files combined to form a single CSV file with 47 columns and 46,686,579 rows. The dataset used for training the DL models contains only 10\% (i.e., 4,668,653 rows) of the original dataset to prevent resource exhaustion. The dataset details are provided in Table \ref{dataset-bi}. The dataset originally contained 47 different features. In order to address the issue of feature dimensionality, the research focused solely on the top 20 features, selecting them based on their importance as generated by a random forest regressor model. The top 20 selected features and feature importance graph are provided in Table \ref{top20feature-tt} and in Fig. \ref{feat-imp}, respectively.  

\begin{table}[t]
  \centering
  \caption{Top 20 selected features \cite{isha1}} \label{top20feature-tt}
  \resizebox{9cm}{!}{%
  \begin{tabular}{llllrr}
    \toprule
    Sr.no & Feature name & Sr.no & Feature name & \multicolumn{1}{l}{Sr.no} & \multicolumn{1}{l}{Feature name} \\
    \midrule
    1     & Srate & 8     & Header\_Length & \multicolumn{1}{l}{15} & \multicolumn{1}{l}{syn\_flag\_number} \\
    2     & Rate  & 9     & flow\_duration & \multicolumn{1}{l}{16} & \multicolumn{1}{l}{rst\_count} \\
    3     & Duration & 10    & Max   & \multicolumn{1}{l}{17} & \multicolumn{1}{l}{UDP} \\
    4     & syn\_count & 11    & HTTP  & \multicolumn{1}{l}{18} & \multicolumn{1}{l}{fin\_count} \\
    5     & Weight & 12    & Protocol Type & \multicolumn{1}{l}{19} & \multicolumn{1}{l}{Variance} \\
    6     & ack\_flag\_number & 13    & urg\_count & \multicolumn{1}{l}{20} & \multicolumn{1}{l}{IAT} \\
    7     & Number & 14    & ack\_count &       &  \\
    \bottomrule
    \end{tabular}%
 } 
\end{table}%

After data preprocessing and feature selection, the proposed CNN-based IDS and LSTM-based IDS were trained by employing an 80-20 train-test split ratio. The dataset considered for the research is imbalanced. So, the research uses \lq stratify\rq as a hyperparameter for distributing all classes equally while train-test split \cite{ssm3}. It helps to handle the data imbalance concern appropriately. So, the model will not be biased toward the majority class.

\add{
\begin{table}[t]
  \centering
  \caption{Dataset Details}
  \label{dataset-bi}
  \begin{tabular}{L{0.9in}C{0.5in}C{0.5in}C{0.5in}}
    \hline
    \textbf{Attack Name} & \textbf{Total Samples} & \textbf{Training Samples} & \textbf{Testing Samples} \\
    \hline
    Dataset\_Binary~\cite{isha1} & 4,668,653 & 3,734,922 & 933,731 \\
    Dataset\_Group~\cite{isha1}  & 4,668,653 & 3,734,921 & 933,732 \\
    Dataset\_Multi~\cite{isha1}  & 4,570,593 & 3,656,475 & 914,118 \\
    \hline
  \end{tabular}
\end{table}
}

\subsection{Architecture of Proposed DL-based IDS}
\subsubsection{Proposed CNN-based IDS}

\begin{figure}[t]
\centerline{\includegraphics[height=3.5cm, width=9cm]{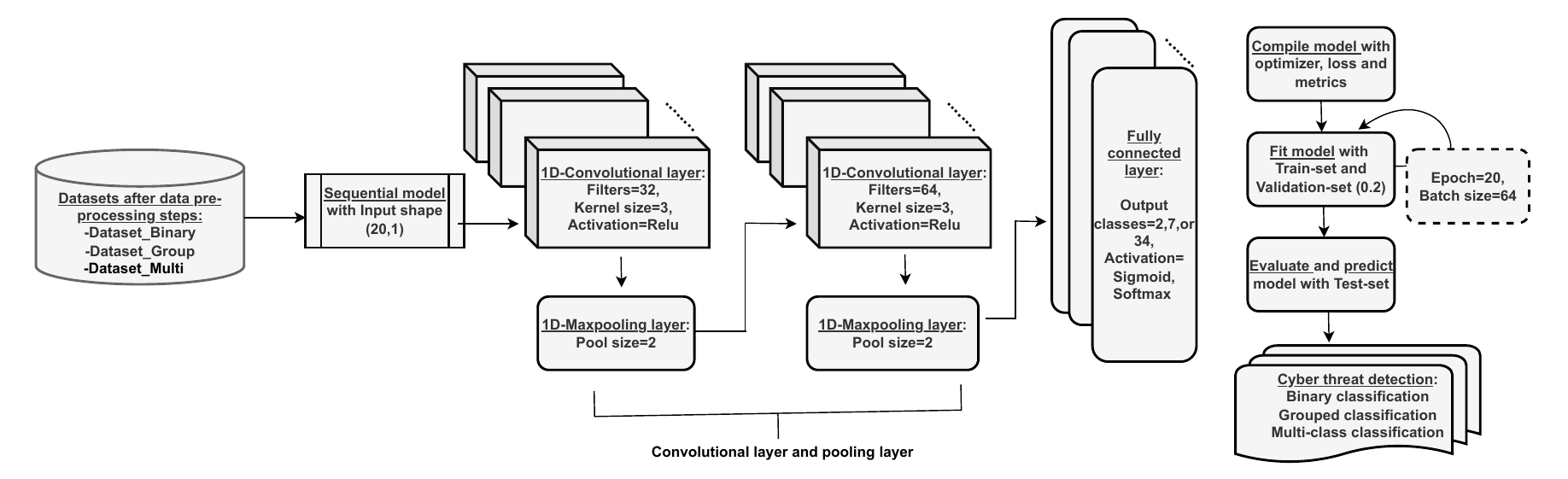}}
\caption{Architecture of proposed CNN-based IDS}
\label{block-cnn-dia}
\end{figure}

CNNs excel in extracting spatial features from data, making them ideal for structured inputs, such as network packet headers \cite{RW11}.
The proposed CNN-based IDS consists of two convolutional layers, two max pooling layers, and a fully connected dense output layer. The first layer is a 1D convolutional layer that has 32 filters with a kernel size of 3 and a \lq relu' activation function, which is followed by a 1D max pooling layer having a pool size of 2. The third layer is a 1D convolutional layer having 64 filters with a kernel size of 3 and a \lq relu' activation function, which is again followed by a max pooling layer of pool size 2. For all three classification, the \lq adam' optimiser is used. The loss function, binary cross-entropy is used for binary and sparse\_categorical\_crossentropy is used for grouped and multi-class classifications. Details of the proposed CNN-based IDS is presented in Fig. \ref{block-cnn-dia} and its hyperparameter settings in \ref{hyperset}.

\subsubsection{Proposed LSTM-based IDS}

\begin{figure}[t]
\centerline{\includegraphics[height=3.5cm, width=9cm]{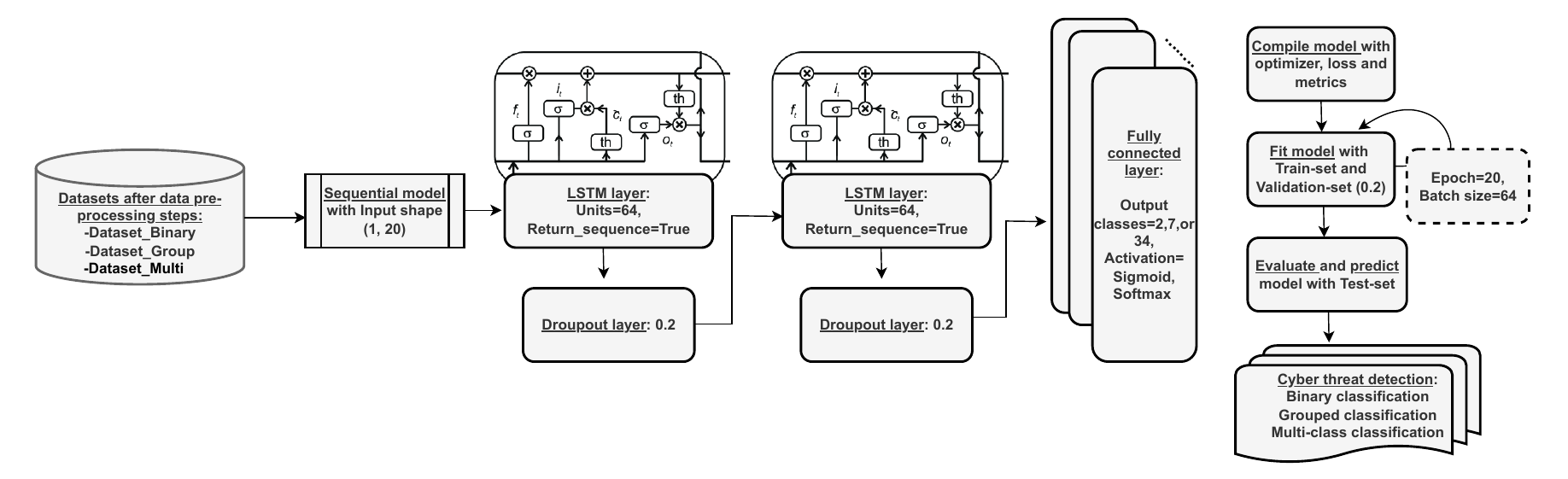}}
\caption{Architecture of proposed LSTM-based IDS}
\label{block-lstm-dia}
\end{figure}

The proposed LSTM-based IDS use two LSTM layers with 64 neurones and two Dropout layers. The \lq sigmoid' activation is used for binary classification and a \lq softmax' activation for grouped and multi-class classification. For all three classification perspectives, the \lq adam' optimiser with a learning rate of 0.0001 and accuracy metrics are used. The details of the proposed LSTM-based IDS is presented in Fig. \ref{block-lstm-dia} and its hyperparameter settings in \ref{hyperset}.

\section{Results and Discussion}\label{sec2}

\subsubsection*{Simulation set-up}

The details of the software and hardware setup is illustrated in Table \ref{setup}. The hyperparameter settings are illustrated in Table \ref{hyperset}. The Keras and TensorFlow Python libraries were used to implement the proposed DL-based IDS.

\begin{table}[t]
  \centering
  \caption{Simulation set-up}\label{setup}
  \resizebox{9cm}{!}{%
    \begin{tabular}{ll}
    \hline
    \multicolumn{2}{l}{Software specification:} \\ \hline
    Software Tool       & Jupyter Notebook 6.5.4 \\ 
    Programming Language & Python 3.11.4, Sklearn library, Matplotlib 3.7.1\\ 
    API Tool            & Keras on TensorFlow 2.14.0 \\ \hline
    \multicolumn{2}{l}{Hardware specification:} \\ \hline
    Processor           & Processor Intel(R) Core(TM) i7-8550U CPU @ 1.80 GHz, Windows 11 64-bit OS \\ 
    RAM                 & 12.0 GB \\ 
    Graphics Card       & Intel(R) UHD Graphics 620 (Integrated Graphics Card) \\ \hline
    \end{tabular}%
  }
\end{table}


\begin{table}[!htbp]
  \centering
  \caption{Hyperparameter settings} 
  \resizebox{9cm}{!}{
    \begin{tabular}{ccll}
      \toprule
      \multicolumn{3}{c}{Hyperparameter settings} & Value \\
      \midrule
      \multicolumn{2}{c}{\multirow{2}[2]{*}{Binary classification}} & Loss       & binary\_crossentropy \\
      \multicolumn{2}{c}{}                                        & Activation & sigmoid \\
      \midrule
      \multicolumn{1}{c}{\multirow{7}[4]{*}{Grouped classification, Multi-class classification}} 
          & \multicolumn{1}{c}{\multirow{3}[2]{*}{Proposed CNN IDS}} 
              & Optimizer    & adam \\
          &                & Activation  & softmax \\
          &                & Loss        & sparse\_categorical\_crossentropy \\
      \cmidrule{2-4} 
          & \multicolumn{1}{c}{\multirow{4}[2]{*}{Proposed LSTM IDS}} 
              & Optimizer    & adam \\
          &                & Activation  & softmax \\
          &                & Learning rate & 0.0001 \\
          &                & Loss        & categorical\_crossentropy \\
      \bottomrule
    \end{tabular}
  } 
  \label{hyperset}
\end{table}

\subsubsection*{Performance metrics}
The research considered standard performance matrices to evaluate the performance of the proposed DL-based IDS \cite{RW5}. They are accuracy, precision, recall, and F1 score.

\subsubsection*{Performance evaluation}
The research investigates the performance of proposed DL-based IDS using the state-of-the-art HetIoT CNN-IDS \cite{ssm1}. The state-of-the-art HetIoT CNN-IDS is a unique CNN model that classifies DDoS attacks in heterogeneous IoT environments. The research employed the same model to assess and compare its performance using the CICIoT2023 dataset. The research changed two parameters: kernel size from 5 to 2 and strides from 2 to 1 when reimplemented the state-of-the-art HetIoT CNN-IDS \cite{ssm1}. The research adopts the same parameters as in \cite{ssm1} to evaluate binary, grouped, and multi-class classification.


\subsection{Binary classification}
\begin{figure}[t]
\centerline{\includegraphics[height=3cm, width=9cm]{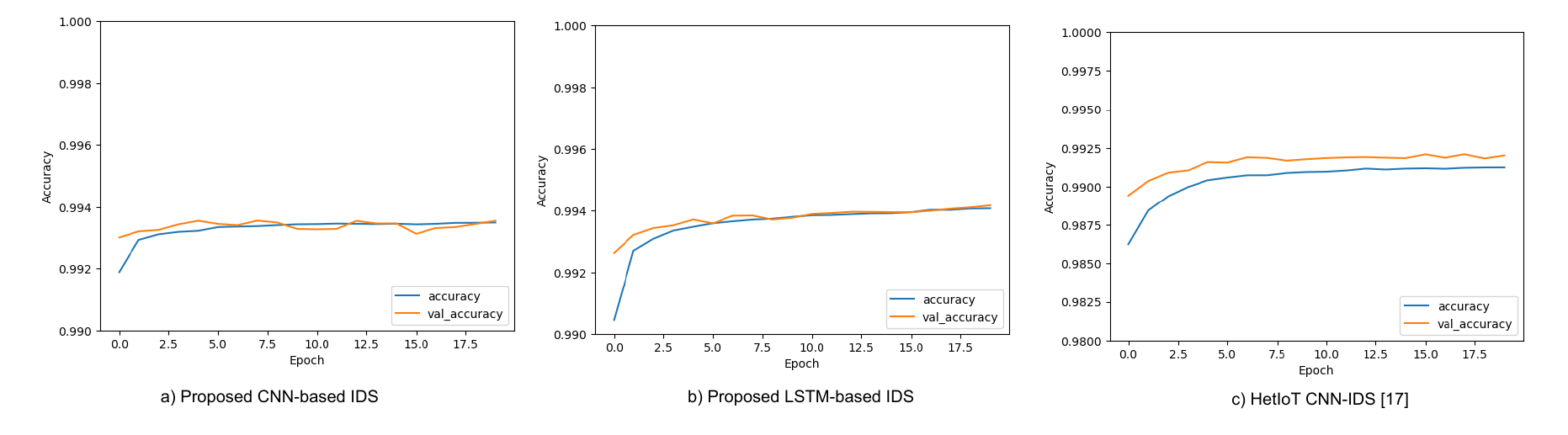}}

\caption{Accuracy of Binary classification}
\label{graph-bi}
\end{figure}

The details of the dataset for binary classification have been provided in Table \ref{dataset-bi}. The hyperparameter settings are presented in the Table \ref{hyperset}. The details of the proposed CNN-based IDS and the proposed LSTM-based IDS are provided in Fig. \ref{block-cnn-dia} and Fig. \ref{block-lstm-dia}, respectively. Fig. \ref{graph-bi}a), \ref{graph-bi}b), and \ref{graph-bi}c) illustrate the training and validation accuracy of the proposed CNN-based IDS, LSTM-based IDS, and state-of-the-art HetIoT CNN-IDS \cite{ssm1} for binary classification across 20 epochs.
The proposed CNN-based IDS achieves an accuracy of 99.34\%, whereas the proposed LSTM-based IDS achieves an accuracy of 99.42\%. The state-of-the-art HetIoT CNN-IDS \cite{ssm1} attained an accuracy of 99.20\%. Fig. \ref{graph-bi}a) shows that the proposed CNN-based IDS performs well without being overfitted or under-fitted. Fig. \ref{graph-bi}b) demonstrates that the proposed LSTM-based IDS performance remains stable after 15 epochs. In comparison to the proposed DL-based IDS, the performance of the state-of-the-art HetIoT CNN-IDS \cite{ssm1} falls short. The graph in Fig. (\ref{graph-bi}c) represents this.

\subsection{Grouped classification}
\begin{figure}[t]
\centerline{\includegraphics[height=3cm, width=9cm]{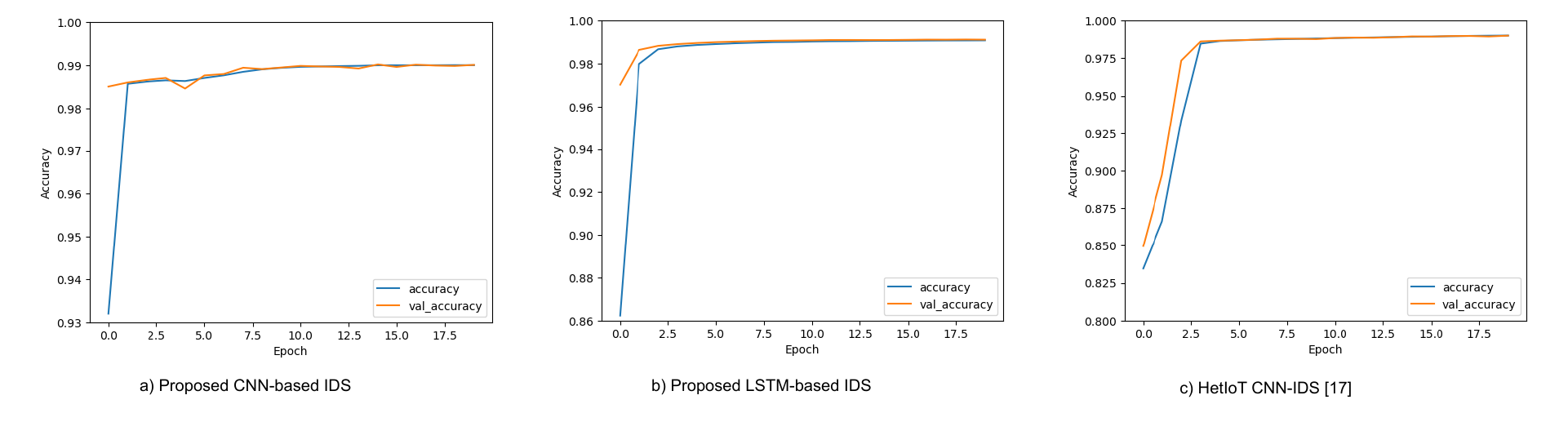}}
\caption{Accuracy of Grouped classification}
\label{graph-grp}
\end{figure}

The details of the dataset for grouped classification have been provided in Table \ref{dataset-grp}. The hyperparameter settings are presented in the Table \ref{hyperset}. The details of the proposed CNN-based IDS and the proposed LSTM-based IDS are provided in Fig. \ref{block-cnn-dia} and Fig. \ref{block-lstm-dia}, respectively. Fig. \ref{graph-grp}a), \ref{graph-grp}b), and \ref{graph-grp}c) illustrate the training and validation accuracy of the proposed CNN-based IDS, LSTM-based IDS, and state-of-the-art HetIoT CNN-IDS \cite{ssm1} for grouped classification across 20 epochs.
The proposed CNN-based IDS obtains a testing accuracy of 99.34\%, whereas the proposed LSTM-based IDS achieves an accuracy of 99.13\%. The state-of-the-art HetIoT CNN-IDS \cite{ssm1} attained an accuracy of 99\%. Fig. \ref{graph-grp}a) demonstrates that the proposed CNN-based IDS performs well after 10 epochs with minor variations. Fig. \ref{graph-grp}b) demonstrates that the proposed LSTM-based IDS performance is better after 8 epochs. The state-of-the-art HetIoT CNN-IDS \cite{ssm1} performance continues to improve as the model is trained and validated with more epochs, shown in Fig. \ref{graph-grp}c).   

\subsection{Multi-class classification}
\begin{figure}[t]
\centerline{\includegraphics[height=3cm, width=9cm]{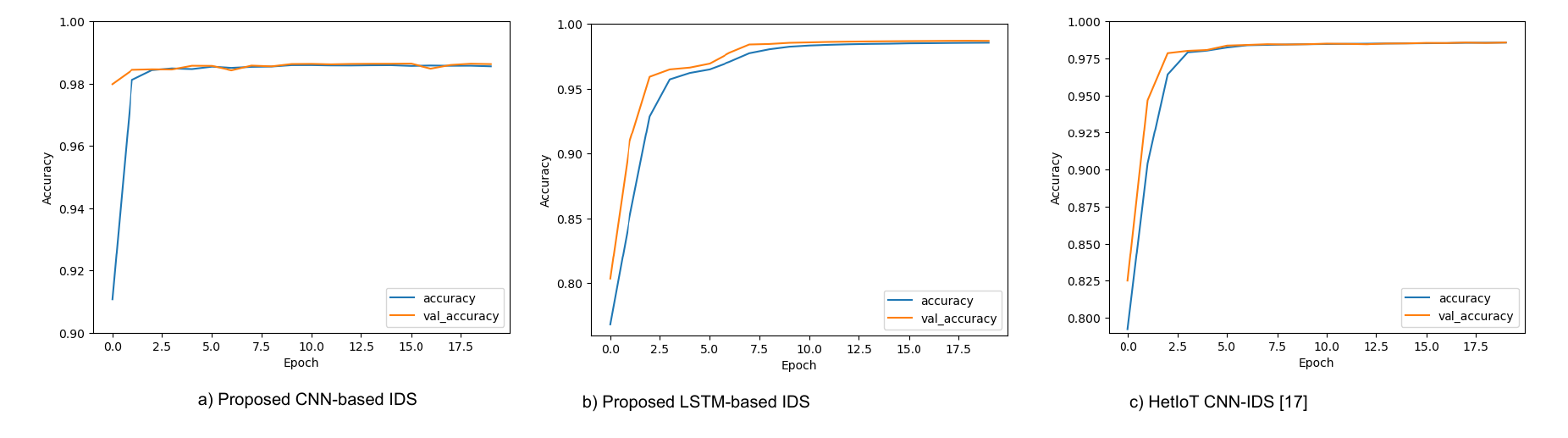}}
\caption{Accuracy of Multi-class classification.}
\label{graph-mul}
\end{figure}

The details of the dataset for multi-class classification have been provided in Table \ref{dataset-mul}. The hyperparameter settings are presented in the Table \ref{hyperset}. The details of the proposed CNN-based IDS and the proposed LSTM-based IDS are provided in Fig. \ref{block-cnn-dia} and Fig. \ref{block-lstm-dia}, respectively. Fig. \ref{graph-mul}a), \ref{graph-mul}b), and \ref{graph-mul}c) illustrate the training and validation accuracy of the proposed CNN-based IDS, LSTM-based IDS, and state-of-the-art HetIoT CNN-IDS \cite{ssm1} for multi-class classification across 20 epochs.
The proposed CNN-based IDS obtains testing accuracy of 98.62\% whereas the proposed LSTM-based IDS achieves accuracy of 98.68\%. The proposed CNN-based IDS performance is shown in Fig. \ref{graph-mul}a). Compared to the proposed CNN-based IDS, the proposed LSTM-based IDS performed slightly differently during training and validation, as shown in Fig. \ref{graph-mul}b). The state-of-the-art HetIoT CNN-IDS \cite{ssm1} performance improves with each epoch, as shown in Fig. \ref{graph-mul}c). During training validation, it increased from 80\% to 97.5\%. The state-of-the-art HetIoT CNN-IDS \cite{ssm1} achieves testing accuracy of 98.55\%.  

\subsection{Summary}
The comparative analysis with the state-of-the-art HetIoT CNN-IDS \cite{ssm1} and the proposed DL-based IDS, namely, the proposed CNN-based IDS and the proposed LSTM-based IDS, has been depicted in Table \ref{summary}. The proposed DL-based intelligent IDS are lightweight in terms of layers and neurones used in the research. The data preprocessing steps, including feature selection and hyperparameter settings, help the research perform well. Compared to the proposed DL-based intelligent IDS, the proposed LSTM-based IDS performs better. The state-of-the-art HetIoT CNN-IDS \cite{ssm1} proposed two separate models for binary and multi-class classifications. However, the research proposed the same IDS for all kinds of classifications, specifically, binary, grouped, and multi-class classifications. Further, the proposed CNN-based IDS performs better compared to the state-of-the-art HetIoT CNN-IDS \cite{ssm1}. Among two DL-based intelligent IDS, the proposed LSTM-based IDS performance is superior.  
 

\begin{table}[t]
  \centering
  \caption{Comparative analysis with state-of-the art HetIoT CNN-IDS \cite{ssm1} and the proposed DL-based IDS}
  \resizebox{9.5cm}{!}{
    \begin{tabular}{lccc}
    \toprule
    \multicolumn{1}{c}{DL-based intelligent IDS} & \multicolumn{3}{c}{Accuracy(\%)} \\
\cmidrule{2-4}          & \multicolumn{1}{p{5.93em}}{Binary classification} & \multicolumn{1}{p{6.43em}}{Gouped classification} & \multicolumn{1}{p{6.855em}}{Multiclass classification} \\
    \midrule
    Proposed CNN-based IDS & 99.34 & 99.02 & 98.62 \\
    Proposed LSTM-based IDS & 99.42 & 99.13 & 98.68 \\
    HetIoT CNN-IDS \cite{ssm1} & 99.2  & 99    & 98.55 \\
    \bottomrule
    \end{tabular}%
}  \label{summary}%
\end{table}%

\section{Conclusion}
In conclusion, the research proposed two lightweight, DL-based intelligent IDS models: the CNN-based IDS and the LSTM-based IDS. The research proposed DL-based intelligent IDS effectively identify and classify binary (2 class) classification, grouped (8 class) classification, and multi-class (34 class) classification. The research also compares the proposed DL-based intelligent IDS techniques with the state-of-the-art HetIoT CNN-IDS. The researchers reimplemented the state-of-the-art HetIoT CNN-IDS on the CICIoT2023 dataset for a fair comparison.  The proposed CNN-based IDS obtained accuracies of 99.34\%, 99.02\%, and 98.62\% for binary, grouped, and multi-class classification. The proposed LSTM-based IDS obtained accuracies of 99.42\%, 99.13\%, and 98.68\% for binary, grouped, and multi-class classification. Future studies could focus on federated learning approaches to ensure network security and data privacy. The research could potentially be extended to include reinforcement learning approaches for large-scale IoT.

\vspace{0.5cm}

\end{document}